\documentclass[11pt]{article}
\def\be {\begin{equation}}
\def\ee {\end{equation}}
\def\bea {\begin{eqnarray}}
\def\eea {\end{eqnarray}}
\begin{document}
\title{{\bf{\Large Quantum tunneling and black hole spectroscopy}}}
\author{
{\bf {\normalsize Rabin Banerjee}$
$\thanks{E-mail: rabin@bose.res.in}},\,
 {\bf {\normalsize Bibhas Ranjan Majhi}$
$\thanks{E-mail: bibhas@bose.res.in}}\\
 {\normalsize S.~N.~Bose National Centre for Basic Sciences,}
\\{\normalsize JD Block, Sector III, Salt Lake, Kolkata-700098, India}
\\[0.3cm]
{\bf {\normalsize Elias C. Vagenas}$
$\thanks{E-mail: evagenas@academyofathens.gr}}\\
 {\normalsize Research Center for Astronomy \& Applied Mathematics,}
\\{\normalsize Academy of Athens,}
\\{\normalsize Soranou Efessiou 4, GR-11527, Athens, Greece}
\\[0.3cm]
}

\maketitle
%
%
%
%
\begin{abstract}
\par\noindent
The entropy-area spectrum of a black hole has been a long-standing and unsolved problem. Based on a recent
methodology introduced by  two of the authors, for the black hole radiation (Hawking effect)
as tunneling effect, we obtain the entropy spectrum of a  black hole. In {\it{Einstein's gravity}},
we show that both entropy and area spectrum are evenly spaced.
But in more general theories (like {\it{Einstein-Gauss-Bonnet gravity}}),
although the entropy spectrum is equispaced, the corresponding area spectrum is not.

\end{abstract}
%
%
%
%
\par\noindent
\section{Introduction}
Since the birth of Einstein's theory of gravitation, black holes have been one of the main
topics that attracted the attention and consumed a big part of the working time of the scientific community.
In particular, the computation of black hole entropy in the semiclassical and furthermore in the quantum regime has been a very
difficult and (in its full extent) unsolved problem that has created a lot of controversy. A closely related issue is the spectrum of this entropy as well as that of the horizon area. This will be our main concern.

     Bekenstein was the first to show
that there is a  lower bound (quantum) in the increase of the area of the black hole horizon when
a neutral (test) particle is absorbed \cite{Bekenstein:1973ur}
\be
(\Delta{A})_{min}=8\pi \l_{pl}^{2}
\label{bekenstein}
\ee
where we use gravitational units, i.e. $G=c=1$, and $\l_{pl}=(G \hbar /c^{3})^{1/2}$ is the Planck length.
Later on, Hod  considered the case of a charged particle assimilated by a Reissner-Nordstr\"om black hole and
derived a smaller bound for the increase of the black hole area \cite{Hod:1999nb}
\be
(\Delta{A})_{min}=4 \l_{pl}^{2}~.
\label{hod}
\ee
At the same time, a new research direction was pursued; namely the derivation of
the area and thus the entropy spectrum of black holes  utilizing  the quasinormal modes of black holes \cite{Hod:1998vk}
\footnote{For some works on this direction see, for instance, \cite{Setare:2003bd} and references therein.}.
In this framework, the result obtained is of the form
\be
(\Delta{A})_{min}=4 \l_{pl}^{2} \ln k
\label{hod1}
\ee
where $k=3$. A similar expression was first put forward by
Bekenstein and Mukhanov \cite{muk} who employed the ``bit counting'' process.
However in that case $k$ is equal to $2$. Such a spectrum can also be derived in the context of quantum
geometrodynamics \cite{vaz}. Furthermore, using this result one can
find the corrections to entropy consistent with Gibbs' paradox \cite{kif}.
%
%
%
%
\par\noindent
Another significant attempt was to fix the Immirzi parameter in the framework of Loop Quantum Gravity  \cite{Dreyer:2002vy}
but it was unsuccessful \cite{Domagala:2004jt}. Furthermore, contrary to Hod's statement for a uniformly spaced
area spectrum of generic Kerr-Newman black holes, it was proven that the area spacing of Kerr black hole
is not equidistant \cite{Setare:2004uu}.
However, a new interpretation for the black hole quasinormal modes was proposed \cite{Maggiore:2007nq}
which rejuvenated the interest in this direction.
In this framework the area spectrum is evenly spaced and the area quantum for the Schwarschild
as well as for the Kerr black hole is given by (\ref{bekenstein}) \cite{Vagenas:2008yi}.
%
%
%
%
While this is in agreement with the old result of Bekenstein, it disagrees with (\ref{hod}).\\
\par\noindent
  In this paper we will use a modified version of the tunneling mechanism \cite{Volovik:1999fc,Paddy, Wilczek,Banerjee:2008ry,Majhi2,Majhi3,Majhi4,Singleton} proposed by two of the authors (RB and BRM) \cite{bm1,bm2}, to derive the entropy-area spectrum of a black hole. In this formalism, a virtual pair of particles is produced just inside the black hole. One member of this pair is trapped inside the black hole while the other member can quantum mechanically tunnel through the horizon. This is ultimately observed at infinity, giving rise to the Hawking flux. Now the uncertainty in the energy of the emitted particle is calculated from a simple quantum mechanical point of view. Then exploiting information theory ({\it{entropy as lack of information}}) and the first law of thermodynamics, we infer that the entropy spectrum is evenly spaced for both {\it{Einstein's gravity}} as well as {\it{Einstein-Gauss-Bonnet gravity}}. Now, since in Einstein gravity, entropy is proportional to horizon area of black hole, the area spectrum is also evenly spaced and the spacing is shown to be exactly identical with one computed by Hod \cite{Hod:1999nb} who studied the assimilation of charged particle by a Reissner-Nordstr\"om black hole. On the contrary, in more general theories like Einstein-Gauss-Bonnet gravity, the entropy is not proportional to the area and therefore area spacing is not equidistant. This also agrees with recent conclusions \cite{Daw,Wei}. \\
\par\noindent
The organization of the paper goes as follows. In section 2, we briefly present the modified tunneling method. In section 3,
we compute the entropy and area spectrum of a black hole solutions of both Einstein gravity and Einstein-Gauss-Bonnet gravity.
Finally, section 4 is devoted to a brief summary of our results and concluding remarks.
\\
%
%
%
%
\par\noindent
\section{The tunneling methodology}
In this section we briefly present the modified tunneling method as developed by two of us \cite{bm1,bm2}. According to the no hair theorem,
{\it{collapse leads to a black hole endowed with mass, charge, angular momentum and no other free parameters}}.
The most general black hole in four dimensional Einstein theory is given by the Kerr-Newman metric \cite{Umetsu:2009ra}.
%
%
%
%
%
%
%
%
%
%

      Now considering complex scalar fields in the Kerr-Newmann black hole background and then substituting the partial wave decomposition of the scalar field in terms of spherical harmonics, it has been shown that near the horizon the action reduces to an effective $2$-dimensional action for a free complex scalar field \cite{Iso:2006ut,Umetsu:2009ra}. 
%
%

   Now from this $2$-dimensional action one can easily derive the equation of motion of the scalar field $\phi$ corresponding to the $l=0$ mode. 
The equation of motion for $\phi$ is given by \cite{Umetsu:2009ra},
\be
\Big[\frac{1}{F(r)}(\partial_t-iA_t)^2 - F(r)\partial^2_r - F'(r)\partial_r\Big]\phi=0 
\label{KN013}
\ee
where
\be
A_t=eV(r)+m\Omega(r);\,\,\, V(r)= -\frac{Qr}{r^2+a^2}, \Omega(r)=-\frac{a}{r^2+a^2},\,\,\,\ a=\frac{J}{M}
\label{new1}
\ee
and 
\be 
F(r) = \frac{\Delta}{r^2+a^2}; \,\,\, \Delta = r^2-2Mr+a^2+Q^2.
\label{f}
\ee
Here $M, J$ and  $Q$ are the mass, angular momentum and electrical charge of the black hole, respectively while $e$ is the charge of the scalar field $\phi$.
Observe that this is just the Klein-Gordon equation for a free scalar field with $U(1)$ gauge field $A_t$ in the following 2-dimensional space-time metric
%
%
\be
ds^2=-F(r)dt^2+\frac{dr^2}{F(r)}~.
\label{1.01}
\ee
%
%
%
%
This shows that near the horizon the theory is dimensionally reduced to a 2-dimensional theory with the metric (\ref{1.01}) \cite{Iso:2006ut,Umetsu:2009ra}.

%
%
%
%
%
%
%
%
%
Now to solve (\ref{KN013}) we employ the standard WKB ansatz for $\phi$
\be
\phi(r,t)=e^{-\frac{i}{\hbar}S(r,t)}.
\label{1.22}
\ee
Then proceeding in a similar way as presented in \cite{bm1,bm2},
we obtain the relations between the modes defined inside and outside of the black hole event horizon:
\bea
&&\phi^{(R)}_{in} = e^{-\frac{\pi\omega}{\hbar \kappa}} \phi^{(R)}_{out}
\label{trans1}
\\
&&\phi^{(L)}_{in} =  \phi^{(L)}_{out}
\label{trans2}
\eea
where $\kappa$ is the surface gravity defined by
\be
\kappa = \frac{1}{2} \frac{d F(r)}{dr} \Big|_{r=r_{+}}~.
\ee
Here ``$in$'' (``$out$'') refer to inside (outside) the event horizon and $L$ ($R$) represent the ingoing (outgoing) mode. In this case $\omega$ is given by the following relation
\be
\omega=E-eV(r_+)-m\Omega(r_+) ~.
\label{work}
\ee
%
%
Here $E$ is the conserved quantity corresponding to a timelike Killing vector. The other variables $V(r_+)$ and $\Omega(r_+)$ are the electric potential and the angular velocity calculated on the horizon.
This $\omega$ is identified as the effective
energy experienced by the particle at asymptotic infinity.
The modes (\ref{trans1},\ref{trans2}) can also be obtained by other approaches \cite{referee}.

    Since the left moving mode travels towards the center of the black hole, its probability to go inside,
as measured by an external observer, is expected to be unity. This is easily verified by computing
\be
P^{(L)}=|\phi^{(L)}_{in}|^2 = |\phi^{(L)}_{out}|^2=1
\label{Krus4}
\ee
where we have used (\ref{trans2}) to recast $\phi^{(L)}_{in}$ in terms of $\phi^{(L)}_{out}$
since measurements are done by an outside observer.
This shows that the left moving (ingoing) mode is trapped inside the black hole, as expected.
On the other hand the right moving mode, i.e. $\phi^{(R)}_{in}$, tunnels through the event horizon
%
%
and its probability, to go outside the horizon, as measured by an external observer is
$P^{(R)}=|\phi^{(R)}_{in}|^2 = |e^{-\frac{\pi\omega}{\hbar \kappa}}\phi^{(R)}_{out}|^2
=e^{-\frac{2\pi\omega}{\hbar \kappa}}$.
%
%

   The same analysis also goes through for a D-dimensional spherically symmetric static black hole
which is a solution for Einstein-Gauss-Bonnet theory \cite{Myers}. This is because the dimensional
reduction technique near the horizon once again tells that the physics can be effectively described by the 2-dimensional
form (\ref{1.01}). Here $F(r)$ is given by
\be
F(r)=1+\frac{r^2}{2\alpha}\Big[1-\Big(1+\frac{4\alpha\bar{\omega}}{r^{D-1}}\Big)^{\frac{1}{2}}\Big]
\label{gauss1}
\ee
with
\bea
\alpha&=&(D-3)(D-4)\alpha_{GB}
\label{gauss2}
\\
\bar{\omega}&=&\frac{16\pi}{(D-2)\Sigma_{D-2}}M
\label{gauss3}
\eea
where $\alpha_{GB}$, $\Sigma_{D-2}$ and $M$ are the coupling constant for the Gauss-Bonnet term in the action,
the volume of unit ($D-2$) sphere and the ADM mass, respectively. Therefore, in the Einstein-Gauss-Bonnet theory
one will obtain the same transformations, namely equations (\ref{trans1}) and (\ref{trans2}),
between the inside and outside modes.
\par\noindent
In the analysis to follow, using the aforementioned transformations,
i.e. equations (\ref{trans1}) and (\ref{trans2}), we will discuss about the spectroscopy
of the entropy and area of black holes.

%
%
%
\par\noindent
\section{Entropy and area spectum}
%
%
In this section we will derive the spectrum for the entropy as well as the area of the black hole defined both in Einstein and Einstein-Gauss-Bonnet gravity. It has already been mentioned that the pair production occurs inside the horizon. The relevant modes are $\phi_{in}^{(L)}$ and $\phi_{in}^{(R)}$.
It has also been shown in the previous section that the left mode is trapped inside the black hole while the right mode can tunnel through the horizon which is observed at asymptotic infinity.
Therefore, the average value of $\omega$ will be computed as
\be
<\omega> = \frac{\displaystyle{\int_0^\infty  \left(\phi^{(R)}_{in}\right)^{*}
\omega  \phi^{(R)}_{in} d\omega}}
{\displaystyle{\int_0^\infty  \left(\phi^{(R)}_{in}\right)^*
\phi^{(R)}_{in} d\omega}} ~.
\label{ref1}
\ee
It should be stressed that the above definition is unique since the pair production occurs inside the black hole and it is the right moving mode that eventually escapes (tunnels) through the horizon.


      To compute this expression it is important to recall that the observer is located outside the event horizon. Therefore it is essential to recast the ``in'' expressions into their corresponding ``out'' expressions using the map (\ref{trans1}) and then perform the integrations.
Consequently, using (\ref{trans1}) in the above we will obtain the average energy of the particle, as seen by the external observer. This is given by,
\bea
<\omega>&=& \frac{\displaystyle{\int_0^\infty e^{-\frac{\pi\omega}{\hbar \kappa}} \left(\phi^{(R)}_{out}\right)^{*}
\omega e^{-\frac{\pi\omega}{\hbar \kappa}} \phi^{(R)}_{out} d\omega}}
{\displaystyle{\int_0^\infty e^{-\frac{\pi\omega}{\hbar \kappa}} \left(\phi^{(R)}_{out}\right)^*
e^{-\frac{\pi\omega}{\hbar \kappa}} \phi^{(R)}_{out} d\omega}}
\nonumber
\\
&=&\frac{\displaystyle{\int_0^\infty  \omega e^{-\beta\omega}d\omega}}
{\displaystyle{\int_0^\infty  e^{-\beta\omega}d\omega}}
\nonumber
\\
&=& \frac{\displaystyle{-\frac{\partial}{\partial\beta}\left(\int_0^\infty  e^{-\beta\omega}d\omega\right)}}
{\displaystyle{\int_0^\infty  e^{-\beta\omega}d\omega}}=\beta^{-1}
\label{spec1}
\eea
where $\beta$ is the inverse Hawking temperature
\be
\beta=\frac{2\pi}{\hbar \kappa}=\frac{1}{T_H}~.
\ee
In a similar way one can compute the average squared energy of the particle detected by
the asymptotic observer
\bea
<\omega^{2}>&=&
\frac{\displaystyle{\int_0^\infty e^{-\frac{\pi\omega}{\hbar \kappa}}\left(\phi^{(R)}_{out}\right)^{*}
\omega^2 e^{-\frac{\pi\omega}{\hbar \kappa}} \phi^{(R)}_{out} d\omega}}
{\displaystyle{\int_0^\infty e^{-\frac{\pi\omega}{\hbar \kappa}}\left(\phi^{(R)}_{out}\right)^{*}
e^{-\frac{\pi\omega}{\hbar \kappa}} \phi^{(R)}_{out}d\omega}}
=\frac{2}{\beta^{2}}~.
\label{spec2}
\eea
Now it is straightforward to evaluate the uncertainty, employing equations (\ref{spec1}) and (\ref{spec2}),
in the detected energy $\omega$
\be
\left(\Delta\omega \right)=\sqrt{<\!\!\omega^{2}\!\!>-<\!\!\omega\!\!>^2}\,=\, \beta^{-1} = T_H
\label{spec3}
\ee
which is nothing but the Hawking temperature $T_{H}$.
Hence the characteristic frequency of the outgoing mode is given by,
\be
\Delta f = \frac{\Delta\omega}{\hbar}=\frac{T_H}{\hbar}.
\label{frequency}
\ee

   Now the uncertainty (\ref{spec3}) in $\omega$ can be seen as the lack of information in energy of the black hole due to the particle emission. This is because $\omega$ is the effective energy defined in (\ref{work}).
Also, since in information theory the entropy is lack of information, 
then the first law of black hole mechanics can be exploited to connect these quantities,
\be
S_{bh}=\int \frac{\Delta \omega}{T_H}.
\label{spec5}
\ee
Substituting the value of $T_H$ from (\ref{frequency}) in the above we obtain
\be
S_{bh} = \frac{1}{\hbar}\int \frac{\Delta\omega}{\Delta f}.
\label{law1}
\ee
Now according to the Bohr-Sommerfeld quantization rule
\be
\int \frac{\Delta\omega}{\Delta f} = n\hbar
\label{law2}
\ee
where $n=1,2,3....$. Hence, combining (\ref{law1}) and (\ref{law2}), we can immediately infer that the entropy is quantized and the spectrum is given by
\be
S_{bh}=n.
\label{entropyspec}
\ee
This shows that the entropy of the black hole is quantized in units of the identity, $\Delta S_{bh} = (n+1) - n = 1$. Thus the corresponding
spectrum is equidistant for both {\it Einstein} as well as {\it Einstein-Gauss-Bonnet} theory.
\par\noindent
Moreover the entropy of a black hole in {\it{Einstein}} theory is given by the Bekenstein-Hawking formula
\be
S_{bh}=\frac{A}{4\l_{pl}^{2}}~.
\label{spec7}
\ee
Consequently, the area of the black hole horizon is also quantized with the area quantum given by,
\be
\Delta A=4\l_{pl}^{2}
\label{spec8}
\ee
implying that the area spectrum is evenly spaced
\be
 A_{n}=4\l_{pl}^{2}\, n\,
\label{spec9}
\ee
with  $n=1,2,3,\ldots $~.
\par\noindent
A couple of comments are in order here. First, the area quantum is universal in the sense that
it is independent of the black hole parameters. This universality was also derived in the
context of the new interpretation of quasinormal moles of black holes \cite{Maggiore:2007nq, Vagenas:2008yi}.
Second, the same value was also obtained earlier by Hod
by considering the Heisenberg uncertainty principle and Schwinger-type charge emission process \cite{Hod:1999nb}.
\par\noindent
On the contrary, in Einstein-Gauss-Bonnet theory, the black hole entropy is given by
\be
S_{bh}=\frac{A}{4}\Big[1+2\alpha\Big(\frac{D-2}{D-4}\Big)\Big(\frac{A}{\Sigma_{D-2}}\Big)^{-\frac{2}{D-2}}\Big]
\label{spec10}
\ee
which shows that entropy is not proportional to area. Therefore in this case the area spacing is not equidistant. This is compatible with recent findings \cite{Daw,Wei}.

\section{Conclusions}
We have calculated the entropy and area spectra of a black hole which is a solution of either Einstein or Einstein-Gauss-Bonnet (EGB) theory.
The computations were pursued in the framework of the tunneling method as reformulated by two of the authors \cite{bm1,bm2}.
In both cases entropy spectrum is equispaced and the quantum of spacing is identical.
Since in Einstein gravity, the entropy is proportional to the horizon area, the spectrum for the corresponding
area is also equally spaced.
The area quantum obtained here is equal to $4\l^{2}_{pl}$. This exactly reproduces the result of
Hod who studied the assimilation of a charged particle by a Reissner-Nordstr\"om black hole \cite{Hod:1999nb}.
In addition, the area quantum $4\l^{2}_{pl}$ is smaller than that
given by Bekenstein for neutral particles \cite{Bekenstein:1973ur} as well as the one computed in the
context of black hole quasinormal modes \cite{Maggiore:2007nq,Vagenas:2008yi}.
\par\noindent
Furthermore, for the computation of the area quantum obtained here, concepts from
statistical physics, quantum mechanics and black hole physics were combined. Therefore, it seems
that the result reached in our analysis is a much better approximation (since a quantum theory of gravity which will give a definite
answer to the quantization of black hole entropy/area is still lacking).
Finally, the equality between our result and that of Hod for the area quantum may be due to the similarity
between the tunneling mechanism and the Schwinger mechanism (for a further discussion on this similarity see \cite{Paddy,Kim}).
On the other hand in EGB gravity, since  entropy is not proportional to area, the spectrum of area is not evenly spaced. Hence,
for EGB gravity, {\it{the notion of the quantum of entropy is more natural than the quantum of area}}.
%
%
However, one should mention that since our calculations are based on a semiclassical approximation,
the spacing obtained here is valid for large values of $n$ and for $s$-wave ($l=0$ mode).
%
%
%
%
%

\end{document}